# Scalable and Deterministic Fabrication of Quantum Emitter Arrays from Hexagonal Boron Nitride

Chi Li[1], Noah Mendelson[1], Ritika Ritika[1], Yong-Liang Chen[1], Zai-Quan Xu[1], Milos Toth[1,2], Igor Aharonovich[1,2*]

1. School of Mathematical and Physical Sciences, Faculty of Science, University of Technology Sydney, Ultimo, New South Wales 2007, Australia
2. ARC Centre of Excellence for Transformative Meta-Optical Systems (TMOS), University of Technology Sydney, Ultimo, New South Wales 2007, Australia

**ABSTRACT**

We demonstrate the fabrication of large-scale arrays of single photon emitters (SPEs) in hexagonal boron nitride (hBN). Bottom-up growth of hBN onto nanoscale arrays of dielectric pillars yields corresponding arrays of hBN emitters at the pillar sites. Statistical analysis shows that the pillar diameter is critical for isolating single defects, and diameters of ~250 nm produce a near-unity yield of a single emitter at each pillar site. Our results constitute a promising route towards spatially-controlled generation of hBN SPEs and provide an effective and efficient method to create large scale SPE arrays. The results pave the way to scalability and high throughput fabrication of SPEs for advanced quantum photonic applications.

**INTRODUCTION**

Solid-state single photon emitters (SPEs) lie at the heart of a wide range of emerging quantum technologies,[1-3] such as quantum memories,[4-6] quantum communications,[7-9] and quantum sensing.[10-13] In recent years, SPEs originating from defects in hexagonal boron nitride (hBN) have been thoroughly investigated as they display outstanding optical properties at room temperature,[14-17] including high brightness,[18] linear polarization,[19] and access to spin states.[20-22] These emitters can occur in commercially-available hBN,[23] or can be engineered by employing various growth techniques such as molecular beam epitaxy,[22, 24] chemical vapor deposition (CVD),[25-28] and metalorganic vapor phase epitaxy.[22] Specifically, CVD methods are attractive due to their low cost, simple experimental setups, short growth times, and the ability to grow hBN over large areas. As a result, hBN CVD protocols have been developed to control material properties such as SPE density and the spectral distribution of SPE zero phonon lines (ZPLs).[27, 29]

In parallel, numerous attempts have been carried out to control the spatial location of SPEs in hBN with high fidelity. This is of paramount importance for SPE-based quantum technologies as it enables practical and scalable integration of quantum emitters with nanophotonic components. Previous studies have explored the preferential occurrence of SPEs at hBN wrinkles[30] and bubbles[31] irradiation of hBN with focused ion beams[32] or electron beams[33] and transfer of hBN onto arrays of pre-fabricated nanopillars.[34, 35] These methods require cumbersome post-processing of the host material, which in most cases significantly damage the material, and lack the scalability that is needed for integrated technologies. While some evidence of SPE localization was observed in most cases, deterministic site-specific generation of narrow-band SPEs and large-scale SPE arrays with a high success rate has remained elusive.

Here we address this challenge by growing hBN directly onto dielectric nanopillars, and show conclusively that under specific conditions, a near-unity yield of SPEs is achieved at the pillar sites. Specifically, we focus on the generation of large-scale emitter arrays through CVD growth onto dielectric materials such as $SiO_2$. We demonstrate an SPE localization precision of ~250 nm, limited by the pillar nanofabrication process. We also demonstrate that nearly 79% of the pillars contain a single emitter, confirmed by autocorrelation measurements, i.e. $g^{(2)}(0)<0.5$. Our results enable accurate placement of hBN SPEs, and hence a promising and scalable route to future integration of hBN SPEs with on-chip nanophotonic components in a bottom-up fashion.

**RESULTS AND DISCUSSION**

Figure 1a schematically illustrates the concept of our approach, where large scale pillar arrays are fabricated from a dielectric substrate ($SiO_2$) and are used as a template for subsequent bottom-up growth of hBN. Pillar diameters and spacings are defined by electron beam lithography using a positive resist and Ni as a hard mask during reactive ion etching (see methods, and Figure S1 for further details). Figure 1b shows an optical image of four $SiO_2$ pillar arrays fabricated with nominal pillar diameters of 300, 500, 700 and 1000 nm. Atomic Force Microscopy (AFM) and Scanning Electron Microscopy (SEM) were used to evaluate the quality of the fabricated pillar structures as well as their depth and lateral size. Figure 1c displays an AFM trace of a SiO$_2$ pillar array with a nominal diameter of 500 nm. The pillars are truncated cones with a height of ~650 nm. Figure 1d shows a tilted (45°) SEM image of the pillar array. The sloped pillar sidewalls seen in the image are a consequence of the employed etch process which yields actual pillar top diameters of 250 nm, 450 nm, 650 nm and 950 nm, respectively (figure S2).

The fabricated pillar arrays were thoroughly cleaned (see methods) and loaded into the CVD chamber for hBN overgrowth. The hBN CVD process was adapted from prior literature on SPE creation during growth on catalytic metal substrates to enable direct growth onto dielectric substrates[29] using ammonia borane and a furnace temperature of 1200° C (see methods). Raman spectroscopy was performed to confirm that the hBN growth was successful under these conditions. Figure 1e shows a spectrum from the ~950 nm pillars. The peak, assigned to the E$_{2g}$ stretching mode of hBN, is centered at 1370.4 cm$^{-1}$ and has a full-width-at-half-maximum of 35.3 cm$^{-1}$.[36] Note that, for other smaller pillars, we observe the same hBN Raman peak position and linewidth.

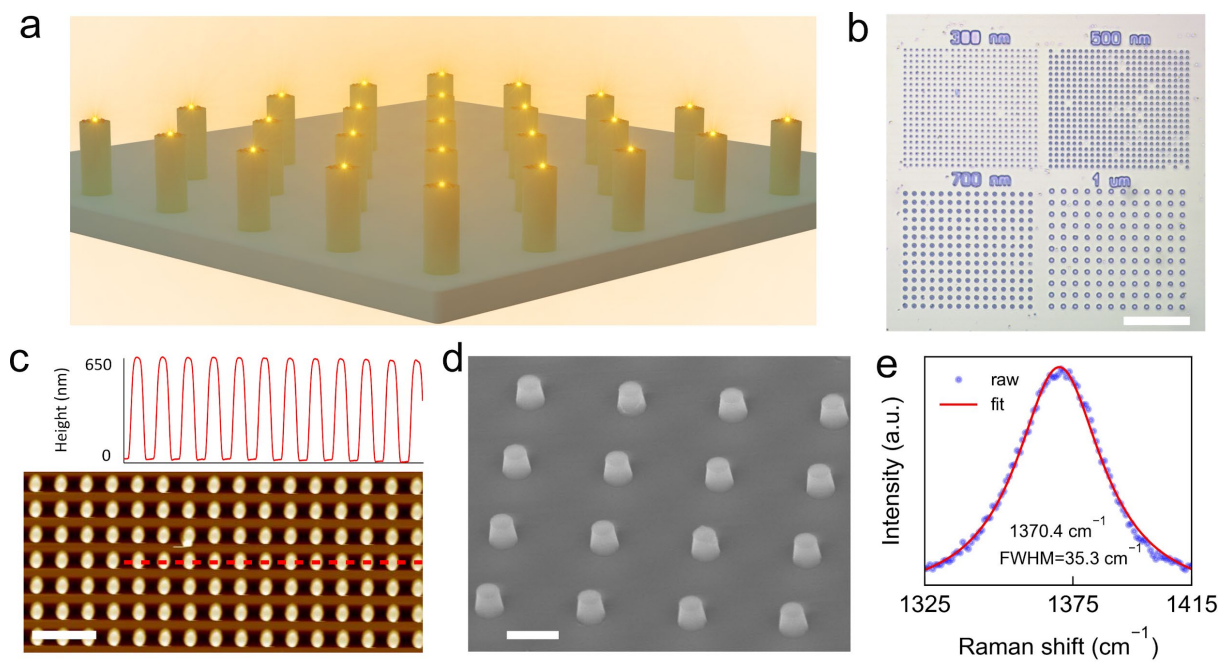

**Figure 1**. *(a) Schematic illustration of single-photon emitters localized by silica pillars. (b) Optical image of four SiO$_2$ pillar arrays with different pillar diameters. The labels – 300 nm, 500 nm, 700 nm and 1 µm – are the nominal pillar diameters. Scale bar: 20 µm. (c) AFM image of pillars in the area labelled "500 nm". The pillar height is ~650 nm. Scale bar: 5um. Upper panel: line profile extracted along the red dashed line in the AFM image. (d) SEM image of pillars in the area labelled "500 nm". The pillar top diameter is ~450 nm and the centre-to-*

*centre distance between the pillars is ~2 um. Scale bar: 1 μm. (e) Raman spectrum of hBN grown on top of the pillars, recorded with a 633 nm laser.*

We next explore the optical properties of hBN grown on SiO$_2$ pillar arrays using confocal photoluminescence (PL) microscopy and wide-field imaging (see methods). Confocal PL was performed at room temperature using a lab-built scanning confocal setup and a 532 nm continuous wave excitation laser (a power of 100 μW was used for all measurements unless noted otherwise) focused on the sample plane (i.e. pillar tops) using a 0.9NA objective. Figure 2a displays a confocal PL scan with an area of ~100 x100 μm, where pillar arrays of four top diameters (~250, ~450, ~650 and ~950 nm) are seen. Bright spots correspond to emission from pillar tops. Note that the spacing of adjacent pillars, i.e. the density, varies for each array. In the 250 nm and 450 nm areas, clear arrays can be resolved in the maps, while they become less evident as the pillar diameter increases due to an overall decrease in fluorescence. Interestingly, this suggests that emission intensity scales with pillar diameter. Figure S3 describes this trend quantitatively, plotting average PL intensity by pixel. We observe emission intensity at a maximum on the ~250 nm pillars, at a minimum for the ~950 nm pillars, and approximately equivalent for the ~450 and ~650 nm pillars. We attribute this trend to increased light collection due to scattering of the SPE fluorescence.

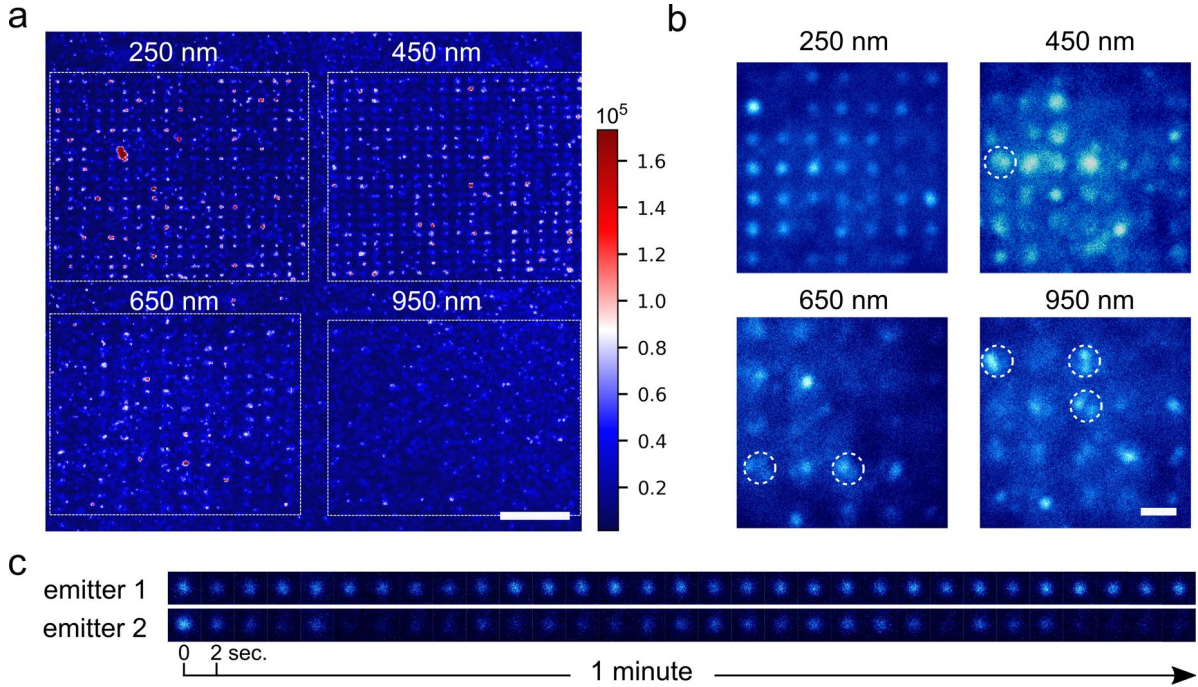

**Figure 2**. *(a) Confocal PL intensity maps of hBN grown on silica pillar arrays. The pillar diameters are 250 nm, 450 nm, 650nm, and 950 nm respectively. Scale bar: 10 μm. The objective is focused on the top of the pillars. The pillar-to-pillar distance varies, e.g. it is 2 μm for 250-nm and 450-nm pillars, 3 μm for 650-nm pillars and 4 μm for the 950-um pillar array.*

*Fluorescence intensities are plotted as photon counts per second. (b) wide-field images obtained from hBN grown on each pillar array. Some examples of pillars containing multiple emitters are indicated by white circles. Scale bar: 2 um. (c) Composited 1-minute recordings of two emitters from 250 nm pillars, showing temporal variations in emission brightness. The time interval is 2 seconds.*

To directly visualize emission centers in the pillar arrays, we analyzed each region by wide-field imaging using a 0.55 NA objective. Figure 2b shows a representative image for each of the four regions. For larger pillar diameters we frequently observe multiple localized emission centers per pillar, each corresponding to SPEs. Some examples are indicated by white dashed circles in Figure 2b. This effect is most clearly seen in the imaged area corresponding to the ~950 nm pillars, where 2-3 distinct SPEs are often seen to occupy individual pillar sites.

In the displayed area corresponding to pillar diameters of ~450 nm we can visualize the structured array more definitively. However, multiple emission centers still appear to occupy each pillar in many cases. Intriguingly, when the pillar diameter is reduced to ~250 nm we can definitively resolve the array geometry, and only a single defect center appears visible on each pillar. Furthermore, SPEs appear to be present on the vast majority of the pillars, suggesting not only the incorporation of a single defect per pillar but also that an SPE array was fabricated with a high success rate.

To further analyze the emission properties of the SPEs on the 250 nm pillars, we select two centers, and record a time resolved acquisition for 1 min, as shown in figure 2c. Emitter 1 appears stable, showing only minor variations in emission intensity, but no obvious blinking events. In contrast, emitter 2 shows both variations in emission intensity as well as blinking events where the emission is not observable, such as at the end of the time series. Overall, we observed most of the emitters displayed only minor variations in emission intensity, and an absence of blinking events, within the displayed 7x7 array of ~250 nm pillars.

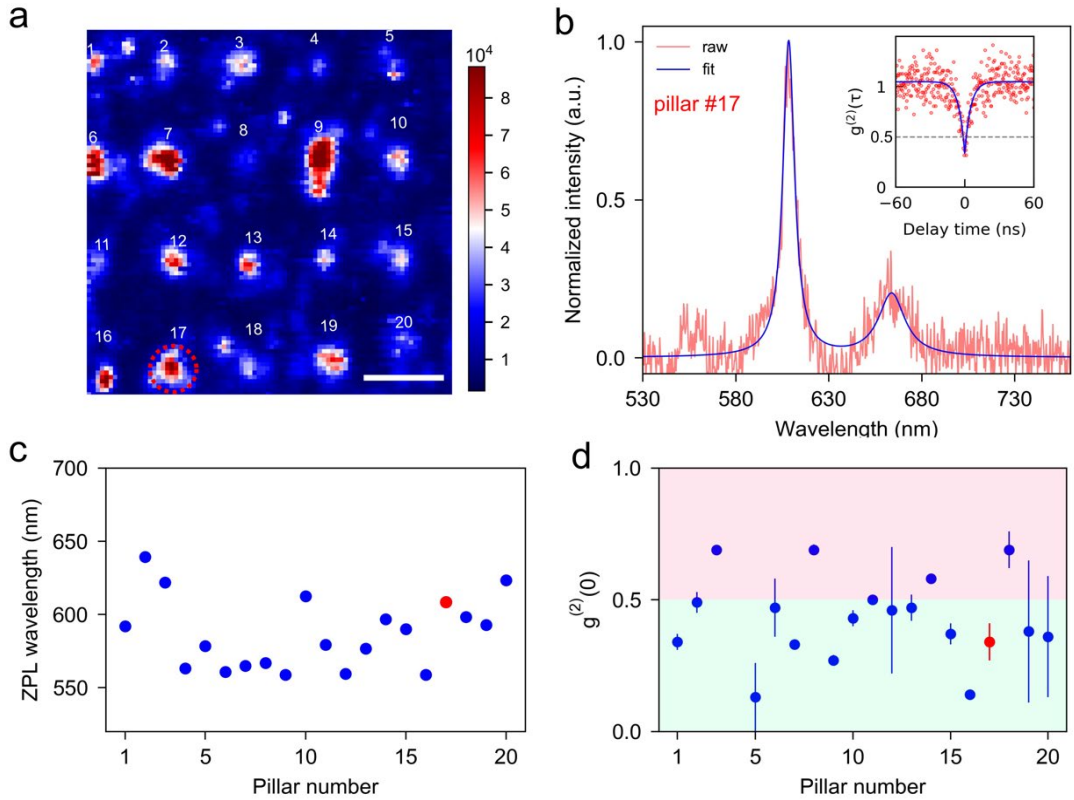

**Figure 3**. *(a) A confocal PL scan of an area including 20 250-nm pillars. Scare bar: 2 μm. Fluorescence intensities are plotted as photon counts per second. Pillar #17 is indicated by a dashed red circle and used as an example in (b). (b) PL spectrum of the single photon emitter responsible for the emission on pillar #17. The two peaks were fitted with Lorentzian functions, and the corresponding ZPL and PSB positions are 608.5 nm and 663.9 nm, respectively. Inset: second-order autocorrelation function $g^{(2)}(\tau)$ curve from the emitter on pillar # 17. (c) and (d) Scatter plots of the ZPL positions (c) and $g^{(2)}(0)$ values (d) extracted from all 20 emitters shown in (a). Error bars in (d) correspond to fitting uncertainties. The red data points in (c) an (d) are from pillar #17. Background subtraction was not applied to the autocorrelation data.*

Given the promising results for the spatially controlled generation of large scale SPE arrays on pillars with a sufficiently small diameter, we then investigated the optical properties of these defects on a single level. Figure 3a displays a confocal scan of a ~10x10 μm area containing 20 pillars with a top diameter of ~250 nm. Bright emission is observed on almost every pillar, consistent with our wide-field imaging results. We subsequently analyzed the spectrum from each pillar top and performed second order autocorrelation measurements, $g^{(2)}(\tau)$, to confirm the quantum nature of the emissions. Figure 3b displays the emission spectrum from pillar number #17, which is representative of that observed from each pillar. The obtained spectrum was fit with two peaks, yielding a ZPL position of ~ 608.5 nm and a phonon sideband (PSB) centered at ~663.9 nm. This corresponds to an energy detuning of ~170 meV between the ZPL and PSB, consistent with prior literature on hBN SPEs.[37] The inset in Figure 3b displays the corresponding $g^{(2)}$ curve, yielding a fitted $g^{(2)}(0)$ value of ~ 0.34, which is below the threshold of $g^{(2)}(0)<0.5$, confirming the quantum nature of this particular emitter. Note that neither background subtraction nor additional spectral filtering was performed for the autocorrelation analysis.

Figure 3c displays a histogram of the fitted ZPL positions recorded from all 20 pillars in Figure 3a, which range from 559-639 nm, with ZPLs lower than 600 nm being the most prevalent, consistent with previous reports on bottom up growth of hBN SPEs.[25, 27] Figure 3d shows the fitted $g^{(2)}(0)$ value recorded from the same set. Without spectral filtering or background subtraction, 79% (15/19) of the pillars exhibit a $g^{(2)}(0) < 0.5$, confirming our technique creates a large-scale array of single photon emitters. Note that the emitter on pillar 4 quenched during the coincidence measurement and is therefore not included in figure 3d. Overall, these results show that direct growth of hBN on nanoscale pillars represents a promising approach to achieving spatial control over SPE incorporation during growth, constituting the first such demonstration via bottom-up growth methods. Additionally, our approach provides a scalable one-step approach, which is not achievable through any top-down techniques, providing key advantages for nanophotonic integration.

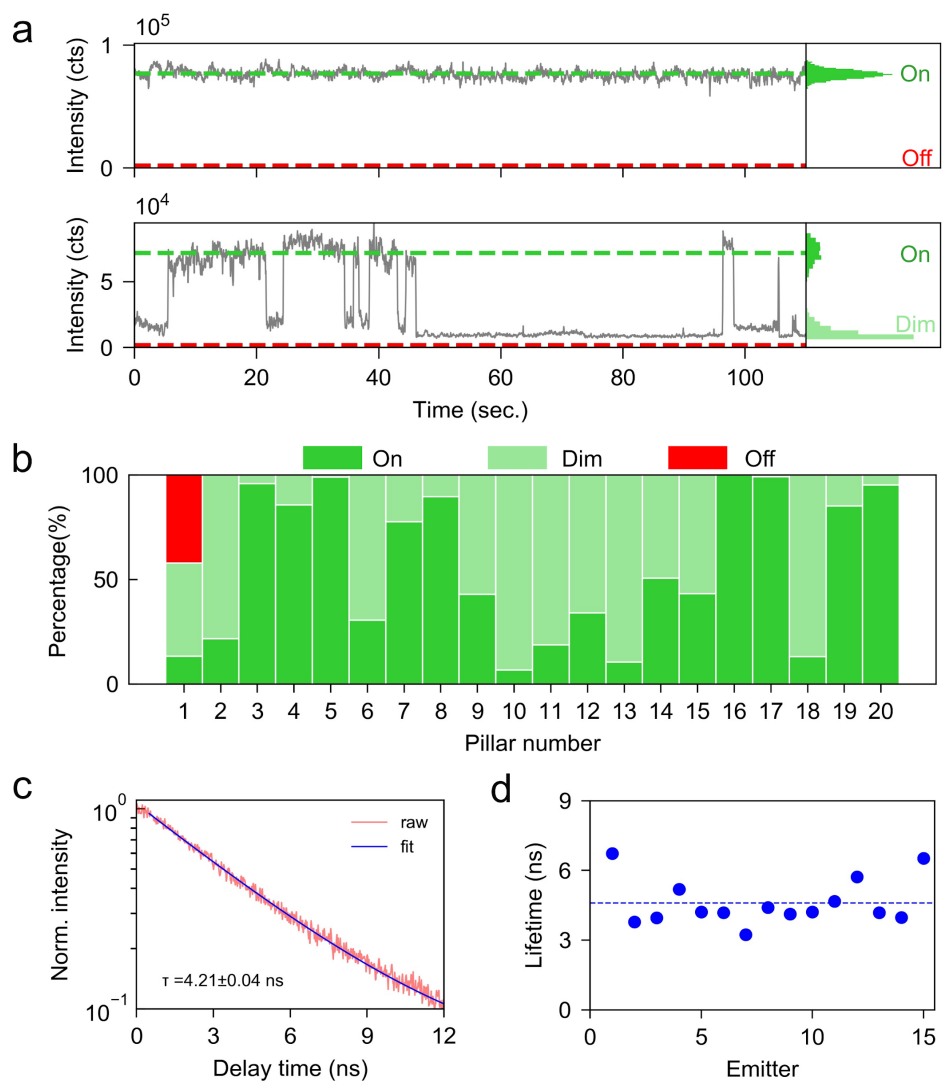

**Figure 4**. *(a) Time resolved fluorescence count rates for two representative SPEs. The top (bottom) trace is from the SPE on pillar 16 (6). Green and red dashed lines denote the emitter on and off states, respectively. The right panels show the corresponding count rate occurrence histograms. The distribution in the bottom histogram contains two modes corresponding to an on state (green) and a distinct dim state (light green) characterized by a count rate that is greater than the background level but much lower than that of the on state. The detector sampling interval is 50 milliseconds. (b) Bar graph summarizing SPE stability from all 20 pillars, expressed as the percentage time spent in the on (green bars), dim (light green bars) and off (red bars) states, respectively. (c) A representative PL decay curve fitted by a single exponential function. The emitter lifetime is 4.21 ± 0.04 ns. (d) Emitter lifetime statistics. The blue dashed line denotes the mean of 4.6 ns. Note that the data where PL decay been plot are not from the previous discussed 20 SPEs.*

Next, we investigate the stability of all 20 SPEs from Figure 3a by recording time-resolved PL intensity traces, and analyzing time spent in different emissive states characterized by distinct PL count rate levels. Traces were recorded for 120 seconds each, and excitation was performed with a 532 nm laser, at a power of 100 µW. For each SPE we observe that the emission rate can be classified into one of three categories at any given time: an on state (corresponding to maximum brightness), an off state (count rates equivalent to the background level of 1000 counts per second), or a dim state (where the emission rate is above background level but reduced substantially relative to the on state). Figure 4a shows two examples of stability traces, with on and off states indicated by dashed green and red lines, respectively. A count rate occurrence histogram is shown on the right of each trace indicating the fraction of time the emitter spent in the on state (by the area under the green peak) and the dim state (light green peak), respectively. The top trace is from the SPE located on pillar #16, and exhibits stable emission behavior, with 100% occupation of the on state. In contrast, the bottom panel corresponds to the SPE on pillar #6, which underwent multiple jumps between the on and dim states, characterized by a bimodal count rate occurrence histogram. The background count rate, corresponding to the off state, is noted in each case by a dashed red line, but is not occupied by either SPE during the acquisition.

Figure 4b summarizes the state occupation statistics for all 20 emitters in the confocal scan area shown in Figure 3a. In each case the relative population of the three states are plotted as a percentage. Only one out of the twenty emitters occupied the off state for any period of time during the measurement. All others fluctuated between on and dim states with roughly half occupying the on state ~ 85% of the time, and the other half rapidly switching between the on and dim states.

Finally, we characterized the emission lifetimes of 15 SPEs incorporated on $SiO_2$ pillars with a diameter of ~250 nm. Figure 4c shows the PL decay trace for one such SPE fitted with a single exponential function which yields a lifetime of ~4.2 ns, consistent with previous literature reports of excited state lifetimes ranging from ~3-6 ns.[19] Figure 4d displays a scatter plot of the lifetimes from all 15 measured SPEs, with the longest lifetime value ~6.7 ns, the shortest of ~3.2 ns, and a mean of ~4.6 ns.

We next explore the formation mechanism of SPE arrays on nanoscale pillars during bottom up growth. Material strain is known to create SPEs for some 2D materials such as the transition metal di-chalcogenides (e.g. $WSe_2$ or $MoS_2$), where exciton recombination can be pinned at local strain maxima.[38, 39] Strain has also been proposed as a method to activate SPEs in hBN,[34] however, the nature of such an activation mechanism is questionable since the energy levels of the emitters are known to be deep states in the wide bandgap of hBN.

To determine whether strain can explain our results, we performed additional control experiments. Specifically, we transferred CVD hBN films (Figure S4) and thin exfoliated hBN flakes (Figure S5) onto pillar arrays. Transfer of CVD hBN films which are known to host a high density of SPEs[27] was performed to determine whether strain activates pre-existing non-luminescent defects. Conversely, transfer of exfoliated hBN flakes which host a low density of SPEs[40] was done to determine whether strain creates new emitters in hBN. For these experiments we used pillars with a slightly reduced heights of ~150 nm and ~350 nm, respectively, to avoid piercing of hBN during transfer. In both cases we observed no evidence for selective formation of SPEs at pillar sites, let alone arrays. This suggests strain is neither a universal nor efficient mechanism for SPE creation or activation in hBN, in contrast to previous work reported in the literature.[34] As a result, strain is unlikely to play a role in the formation of the SPE arrays that we report in the present work.

Having ruled out strain as a mechanism, we discuss two alternatives to explain our results. The first is that we average one SPE per pillar because the pillar density and diameter are optimal for the density of emitters formed during hBN CVD onto a $SiO_2$ substrate. This mechanism is therefore simply a spatial sampling process that isolates particular emitters in the hBN film. It accounts for a number of our results – for instance, the observation that multiple emitters are often present on pillars with larger diameters, whilst a reduction of pillar width leads to an average of a single emitter per pillar. However, the near-unity incorporation rate observed for pillars of ~250 nm is unexpected given the probabilistic nature of this mechanism. Hence, a second more subtle characteristic of the growth process on nanostructured surfaces is expected to play a role. A likely possibility is that the hBN nucleates preferentially at pillar edge sites, and thus affects defect formation. As a consequence, sufficient reduction of the pillar diameter limits nucleation sites, and leads to an average of one SPE per pillar. This mechanism explains the observed relationship between SPE incorporation and pillar diameter, but also more clearly clarifies the near-unity incorporation of a single emitter per pillar for pillar diameters of ~250 nm. Modulating the number of nucleation sites, and thus the number of defects incorporated during growth, provides an elegant and adaptive mechanism for spatially-controlled SPE incorporation during growth.

**CONCLUSION**

In summary, we have demonstrated the deterministic formation of quantum emitters with spatial control by growing hBN directly onto SiO$_2$ pillars. The pillars diameter is critical for isolating a single emitter per pillar, and diameters of ~250 nm yield a near-unity success rate of emitter creation; specifically, ~80% of the pillars exhibit a $g^{(2)}(0)$ value of less than 0.5, characteristic of single photon emission. While the underlying mechanism for the spatial localization remains unknown, our results on transferred hBN flakes convincingly prove that strain alone is not sufficient for the generation or activation of hBN emitters. Our results constitute an advance in the controlled engineering of arrays of solid-state quantum light sources, and specifically quantum emitters in hexagonal boron nitride. These techniques provide a vital step towards integration of these sources with photonic devices, and with optoelectronic circuitry to achieve a room temperature electrically driven SPE in 2D materials.

**METHODS AND MATERIALS**

*hBN growth on dielectric substrates.* The pillar substrates were immersed in hot Piranha solution for 1 hour to remove organic residuals. This was followed by a triple rinse of the substrate in deionized water. A quartz container containing the clean substrates was loaded into the CVD chamber immediately after a 10-minute UV-Ozone clean. At the same time, 50 mg of ammonia borane powder (Sigma-Aldrich) was loaded into a metal crucible and attached to the system. The system was pumped down to 2 mTorr overnight and then annealed at 1200°C for 1 hour under flowing Ar/H$_2$ (300 sccm) to remove trace amounts of surface contaminants. The precursor was heated independently at 95°C to initiate evaporation. Growth took place for 1 hour after the evaporation began, the chamber was kept at 1200°C with an Ar/H$_2$ flow rate of 50 sccm and a pressure of 2 Torr. The system was cooled to room temperature after growth.

*Photonic measurements.* The main PL experiments were performed with our lab-built confocal system. Same as our early reports, briefly, a 532nm CW laser (Gem 532TM, Laser Quantum Ltd.) was used as the excitation source. The laser was reflected by a 532-nm long pass dichroic mirror into a 100x magnification objective (0.9 NA). Scanning was performed using a scanning mirror. A 550-nm long pass filter was placed after the dichroic mirror and before either the spectrometer (Princeton Instruments, Inc.) or two single-photon avalanche diodes (Excelitas Technologies).
For time-resolved PL spectroscopy, a 512 nm pulsed laser (PiL051XTM, Advanced Laser Diode System GmbH) with a 20 MHz repetition frequency was used as the excitation source. The PL emission was synchronized by a correlator (PicoHarp300, PicoQuant).
For wide-field imaging, a 532 nm CW laser with a power of 8 mW was used for excitation. To achieve a large excitation area, the laser was then focused to the Fourier plane of a 40x objective (0.55 NA) to achieve collimated illumination. The allowed field-of-view is approximately 40 x 40 μm$^2$. A 568-nm long pass filter was placed before an EMCCD (Andor iXon Ultra 888).

*Atomic force microscopy.* Pillar heights and the thickness of exfoliated hBN flakes were measured by non-contact AFM (PARK XE7).

*Raman spectroscopy.* Raman spectroscopy was carried out by a confocal Raman system (Renishaw). The excitation source was a 633 nm laser.

## ACKNOWLEDGEMENTS

The authors thank Milad Nonahal for assistance with RIE, and Minh Nguyen for assistance with wide-field imaging. The authors also thank Johannes Forch and Simon White for useful discussions. We acknowledge the Australian Research Council (DP180100077, CE200100010, DP190101058) and the Asian Office of Aerospace Research and Development (FA2386-20-1-4014) for the financial support.